\pdfoutput=1
\documentclass[conference,a4paper]{IEEEtran}
\IEEEoverridecommandlockouts
\usepackage{tikz}
\usetikzlibrary{positioning}
\usepackage{cite}
\usepackage{amsmath,amssymb,amsfonts}
\usepackage{algorithmic}
\usepackage{graphicx}
\usepackage{textcomp}
\usepackage{xcolor}
\def\BibTeX{{\rm B\kern-.05em{\sc i\kern-.025em b}\kern-.08em
    T\kern-.1667em\lower.7ex\hbox{E}\kern-.125emX}}
\begin{document}

\def\correct<#1>{\textbf{* #1 *}}
\def\citat{\textbf{* citation *}}

\title{Can we still use PEAQ? A Performance Analysis of the ITU Standard for the Objective Assessment of Perceived Audio Quality}

\author{\IEEEauthorblockN{Pablo M. Delgado }
\IEEEauthorblockA{
\textit{
\protect\thanks{$^{\ddagger}$ A joint institution between the Friedrich-Alexander Universität Erlangen-Nürnberg (FAU) and Fraunhofer IIS, Germany.}
International Audio Laboratories Erlangen $^{\ddagger}$}\\
Erlangen, Germany \\
pablo.delgado@audiolabs-erlangen.de}
\and
\IEEEauthorblockN{Jürgen Herre}
\IEEEauthorblockA{
\textit{International Audio Laboratories Erlangen $^{\ddagger}$}\\
Erlangen, Germany \\
juergen.herre@audiolabs-erlangen.de}
}

\IEEEoverridecommandlockouts 

\maketitle

\begin{abstract}
The Perceptual Evaluation of Audio Quality (PEAQ) method as described in the International Telecommunication Union (ITU) recommendation ITU-R BS.1387 has been widely used for computationally estimating the quality of perceptually coded audio signals without the need for extensive subjective listening tests. However, many reports have highlighted clear limitations of the scheme after the end of its standardization, particularly involving signals coded with newer technologies such as bandwidth extension or parametric multi-channel coding. Until now, no other method for measuring the quality of both speech and audio signals has been standardized by the ITU. Therefore, a further investigation of the causes for these limitations would be beneficial to a possible update of said scheme. Our experimental results indicate that the performance of PEAQ's model of disturbance loudness is still as good as (and sometimes superior to) other state-of-the-art objective measures, albeit with varying performance depending on the type of degraded signal content (i.e. speech or music). This finding evidences the need for an improved cognitive model. In addition, results indicate that an updated mapping of Model Output Values (MOVs) to PEAQ's Distortion Index (DI) based on newer training data can greatly improve performance. Finally, some suggestions for the improvement of PEAQ are provided based on the reported results and comparison to other systems.

\end{abstract}

\begin{IEEEkeywords}
PEAQ, ViSQOL, PEMO-Q, objective quality assessment, audio quality, speech quality, auditory model, audio coding.
\end{IEEEkeywords}

\section{Introduction}

Efforts initiated in 1994 by the ITU-R to identify and recommend a method for the objective measurement of perceived audio quality culminated in 2001 with recommendation BS.1387 \cite{PEAQ}, most commonly known as the Perceptual Evaluation of Audio Quality (PEAQ) method. This method is based on generally accepted psychoacoustic principles and has successfully been adopted by the perceptual audio codec development and the broadcasting industries \cite{PoctaQuality}.
  
\begin{figure}
\includegraphics[trim={0cm 0 0cm 0},clip, width=0.5\textwidth]{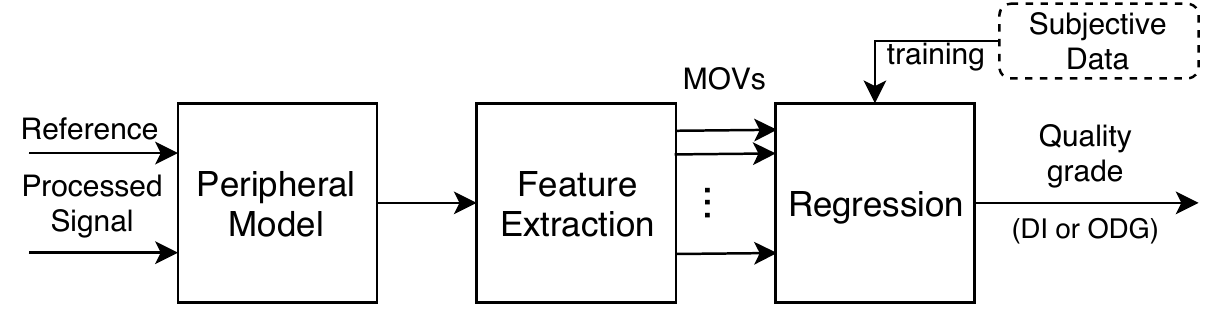}
\caption{High level representation of the PEAQ method \cite{ThiedePEAQ}.}
\label{PEAQ_Model}
\end{figure}    
  
Like numerous other objective audio quality assessment systems, PEAQ takes as inputs a possibly degraded signal (e.g. by bit rate reduction strategies in audio coding) and an unprocessed reference and then compares them in a perceptual domain (i.e. the internal representations of said signals \cite{beerends1992a}) aided by a peripheral ear model and a feature extraction stage. The method then calculates one or more objective degradation indices related to some aspect of perceived quality degradation (Fig. \ref{PEAQ_Model}). In PEAQ, these unidimensional indices (i.e. features) are termed Model Output Values (MOVs) \cite{PEAQ}. The features are mapped to a perceived overall quality scale using regression models trained with subjective data. In the case of PEAQ, several MOVs are combined and mapped to an objective quality scale termed Objective Difference Grade (ODG) using an Artificial Neural Network (ANN). In addition to the ODG output, a Distortion Index (DI) is derived by removing the output layer of the ANN ODG mapping \cite{ThiedePEAQ} to provide a scale-independent condensed degradation index. Both the MOV combination to a DI and the mapping to an ODG scale are carried out using training data from subjective experiments with Subjective Difference Grade (SDG) scores as defined in recommendation ITU-R BS.1116 \cite{BS1116}.

Limitations of objective quality assessment systems can be related to missing models of psychoacoustic phenomena, unsuitable parameter tuning or errors in the degradation index-to-quality mapping. Some solutions to PEAQ limitations are introduced in PEMO-Q \cite{PEMOQ} with the introduction of a modulation filter bank and a degradation measure that considers instantaneous values instead of averages. Various PEAQ extensions for better stereo and multi-channel audio prediction were proposed with improved performance (e.g. \cite{kmpf2010standardization}). However, these extensions did not address existing problems related to timbral quality estimation.

Predicting perceived audio quality degradation by performing comparisons in the (partial) loudness domain \cite{ThiedePEAQ, beerends1992a} has proven to be a successful approach incorporated in many standard objective quality measurement systems \cite{POLQAcite, PEAQ}. The PEAQ MOVs based on the partial loudness domain model is meant only to reliably work within the realm of small disturbances near the masked threshold in the distorted signals. The Perceptual Objective Listener Quality Assessment (POLQA \cite{POLQAcite}) method extended the disturbance loudness model to consider larger disturbances as well. 

Regarding the cognitive modeling of distortion perception, the asymmetric weighting of added and missing time frequency components between degraded and reference internal representations is also widely used \cite{POLQAcite, PEAQ, PEMOQ}. It is generally assumed that added disturbances in the degraded signal will contribute more importantly to the perceived distortion than missing components. However, there has been only sparse literature analyzing whether this relationship holds the same for all kinds of signals. 

Most previous performance comparisons of PEAQ use either the DI or ODG outputs (e.g. \cite{PoctaQuality}, \cite{ViSQOLAudio}), which are dependent on the original training data. This fact can be problematic as the mapping can interfere with the performance evaluation of the intrinsic perceptual model. Alternatively, the work presented in \cite{torcoli2018comparing} showed that separate MOVs of PEAQ have good prediction ability for different types of distortions, including those degradations related to newer audio coding technologies \cite{USACCodec_short} like tonality mismatch or unmasked noise. 

\section{Method}

\subsection{PEAQ's disturbance loudness model evaluation}
\label{PEAQMOV}

As a first objective, we want to evaluate the performance of PEAQ's partial loudness disturbance model in the context of the perceived degradations caused by newer audio codecs in a wide quality range. Secondly, we want to investigate whether the relationship between the different types of disturbances due to added or missing time/frequency components is dependent on signal content type (i.e. audio or speech).

For this purpose, we consider the three internal comparisons performed by PEAQ's advanced version in the disturbance loudness domain separately. These three internal comparisons should describe possible degradations due to linear distortions ($AvgLinDist_A$), due to additive disturbances ($RmsNoiseLoudness_A$) and due to missing components ($RmsMissingComponents$)\cite{PEAQ} in the degraded signal under test in comparison to the reference. From these three comparisons, only two MOVs are fed into the final regression stage: $AvgLinDist_A$ and 
\begin{multline}
RmsNoiseLoudAsym_A = RmsNoiseLoudness_A \\ +0.5*RmsMissingComponents.
\end{multline}

The independent analysis of added versus missing components requires that we analyze all three comparisons separately and not the combined MOV values. For this, we use our own \cite{PEAQ} MATLAB implementation of PEAQ\footnote{Compared for reproducibility against MOVs of the PEAQ OPERA 3.5 distribution with the same settings as \cite{ViSQOLAudio}. The RMSE values of all MOV differences represent at most 5\% of the SD of said MOVs for the used database.} and consider $RmsNoiseLoudness_A$ and $RmsMissingComponents$ as two different MOVs instead of $RmsNoiseLoudAsym_A$. 

\subsection{Subjective audio quality database}
\label{DB}

We use the Unified Speech and Audio Coding Verification Test 1 Database (USAC VT1) \cite{USACdatabase} as ground truth for performance testing. The database contains newer generation audio codecs in comparison to those used in \cite{PEAQ}, which contain newer coding technologies \cite{USACCodec_short} that cause artifacts known to challenge current objective quality measurement systems.

The USAC VT1 database contains a total of 27 reference mono audio excerpts (samples) corresponding to different sample types: 9 music-only samples, 9 speech-only samples and 9 mixed speech and music samples, which makes it convenient for separately studying the performance of the objective measurement systems for speech and music signals.  The database includes of a total of 12 treatments of the samples including $3.5$ and $7.5$ kHz anchors plus three perceptual audio codecs (AMR-WB+, HE-AAC v2 and USAC) at a wide range of bit rates (8kbps to 24kbps) as degraded signals (i.e. test items).  

The subjective quality assessment test was carried out following the MUltiple Stimuli with Hidden Rreference and Anchor (MUSHRA) method \cite{MUSHRA} using 62 listeners from 13 different test sites with previous training and post-screening. Gender ratio of the listeners was not reported. The subjective scores ranging from 0 points (bad quality) to 100 points (excellent quality) are pooled into mean quality MUSHRA scores and 95\% confidence intervals for each sample/treatment combination over all listeners. The subjective scores gave an average of 37 points (SD $16.79$) for the worst quality codec and 77 points (SD $14.35$) for the best -sounding codec, so the quality can be considered to contain a wide audio quality range and not just small disturbances.  In total, 324 condensed data points are used, which will be used as training and validation data as explained in the following paragraphs.

\subsection{MOV to objective quality score mapping}
\label{MARS}

Most of the objective measurement systems use some kind of regression procedure to map one or more objective quality degradation measures to quality scores by using subjective test data \cite{ViSQOLAudio, POLQAcite,PEAQ, PEMOQ}. As is the case with PEAQ's DI, most systems also provide outputs representing a generalization of the objective quality without any mapping to a specific quality scale (i.e. the degradation indices). They should monotonically map quality scores, but should not depend on scales or scale anchors. As explained in the Introduction, PEAQ's DI is scale-independent, but still depends on the original subjective training data \cite{ThiedePEAQ}.

This work will require the remapping of the degradation measures (excluding each system's individual quality scale outputs from previous training) to the subjective data in order to abstract the system performance from the regression stage to focus on the underlying signal processing model.

Our mapping procedure is carried out using a MATLAB implementation of the Multivariate Adaptive Regression Splines (MARS) model \cite{Jekabsons_areslab}, which is a piecewise-linear and piecewise-cubic regression model. The model performs a Generalized Cross Validation (GCV) on the data used for building the model which approximates the leave-one-out cross-validation method. This technique is known to be robust to overfitting when a limited amount of training data is available.  

\subsection{Bootstraping and inference statistics}
\label{Bootstrap}
One question related to the problem of mapping degradation measures to a quality scale is --on one hand-- whether a given mapping will reliably be able to predict the audio quality outside of the realm of the available training data. On the other hand, there is the need for separately evaluating the feature extraction stages (which need to show a strong correlation with some aspect of subjective quality) from the mapping/regression stages.  Our approach to tackle this problem with the available training data is to use a bootstrap technique using Monte Carlo simulations. A similar approach has also made its way into the latest revision of the MUSHRA recommendation \cite{MUSHRA} for subjective quality data analysis.

For each realization of the experiment, we divide the data points into two disjoint sets: the model-building data (training and cross-validation, 80\% of the items) and test data (remaining 20\%). The model-building dataset is used to train and cross-validate a MARS model that maps the different degradation measures (i.e. MOVs) to a single objective quality score. The trained model remains unchanged and is in turn evaluated in its ability to predict the subjective quality scores of the test dataset from the MOVs calculated for the corresponding test items. This procedure is repeated $N=2000$ times, each realization randomly samples (with replacement) the data points belonging to training and test datasets respectively. 

Resampling the data presents advantages in generalization when small amounts of data are available.  Additionally, problems with the regularization and normalization of scores from different experiments/laboratories are less severe as the data was likely gathered with the same testing procedure and initial conditions. The bootstrap method allows us -- to a certain degree -- to focus on the performance behind the feature extraction while still using real world data for validation. Taking mean system performance figures (see \ref{PerfMetrics}) over a sufficient set of trained regression models, we are effectively decoupling the influence of the training data on the performance measurement of feature extraction stage. However, for a general final system performance analysis, as many and diverse subjective test databases as possible should be used.

\subsection{System performance metrics}
\label{PerfMetrics}

System performance metrics are always calculated assessing how well the system's objective measurement output predicts subjective scores that have not been used for training (i.e. from the test dataset). This work uses Pearson's ($R_p$) and Spearman's ($R_s$) correlation between objective and subjective scores and Absolute Error Score (AES) as measures. The AES can be seen as an extension of the usual RMS error that also considers weighting by the confidence intervals (CI) of the mean subjective score \cite{PEAQ}.

For the Monte Carlo simulations, point estimates of the mean performance metric values (with an additional $m$ subscript) and their associated 95\% CI are provided. Confidence intervals from the Monte Carlo simulations enable us to provide a notion of the significance of the performance margins among systems in a direct way.

\section{Results and Discussion}

Table \ref{tab1} shows the overall baseline system performance for the used database. This overall performance was obtained by comparing each system's output objective scores against the totality of the subjective scores of the database without any bootstraping or retraining. The objective scores PEAQ ODG and PEAQ DI represent the output of the advanced version of our implementation of PEAQ's Objective Difference Grade (ODG) and Distortion Index (DI) respectively. PEMO-Q ODG and PEMO-Q PSMt are the ODG and similarity outputs of the PEMO-Q implementation found in \cite{PEASS}. ViSQOLA MOS and ViSQOLA NSIM are the Mean Opinion Score (MOS) and similarity outputs respectively of the ViSQOL Audio impementation found in \cite{Visqol_soft}. All the model outputs related to ODG or MOS contain a built-in regression stage trained with listening test subjective scores obtained with different test methodologies. Outputs ViSQOLA NSIM, PEMO-Q PSMt and PEAQ DI are the degradation indices, of which only PEAQ DI still depends on built-in training data, as explained in section \ref{MARS}. 

All objective measures except those of ViSQOL audio show a weak correlation against subjective scores. Still, a higher Spearman correlation value $R_s$ --which measures rank preservation --  indicates that the systems could be used in the context of one-to-one comparisons (e.g. testing effects of new coding tools within the same audio codec).

ViSQOL Audio's degradation index NSIM is clearly the best performing feature. However, as PEAQ DI --also a degradation index-- is still dependent on the original training data, it might not represent the true performance of the underlying perceptual model. We present a performance analysis of different MOVs without any built-in training in Table \ref{tab2}.

\begin{table}[htbp]
\caption{Baseline System Performance}
\begin{center}
\begin{tabular}{|c|c|c|c|}
\hline
\textbf{Objective}&\multicolumn{3}{|c|}{\textbf{System Performance}} \\
\cline{2-4} 
\textbf{Measure} & \textbf{\textit{$R_p$}}& \textbf{\textit{$R_s$}}& \textbf{\textit{AES}} \\
\hline
\hline
PEAQ ODG		&	0.65		&	0.7	&	6.88 \\
\hline
PEAQ DI			&	0.65		&	0.7	&  2.85 \\
\hline
PEMO-Q ODG	&	0.65		&	0.71	&	3.05	\\
\hline
PEMO-Q PSMt	&	0.64 	&	0.71	&	3.56	\\
\hline
ViSQOLA MOS	& 0.76		&	0.83	&	2.84 \\
\hline
ViSQOLA NSIM	&	0.83		& 0.83	& 2.47  \\ 	
\hline
\end{tabular}
\label{tab1}
\end{center}
\end{table}

Table \ref{tab2} shows the average system performance measures obtained by the bootstrap method from Section \ref{Bootstrap} for combinations of outputs of the evaluated objective measurement systems. The table is also divided into performance measures for music-only items, speech-only items and all items together. Multiple realizations of MOV-to-quality-score mappings of the shown objective measures were trained and tested with disjoint sets of data points. Performance measures are calculated exclusively on the test data for each realization of the experiment. The amount of iterations $N$ has proven to be sufficient so that the 95\% CI do not overlap. Therefore, the difference in performance figures support significant effect sizes. Additionally, the rank order in performance of the objective measures that have been evaluated in Table \ref{tab1} and Table \ref{tab2} is preserved, which speaks for the reliability of the bootstrap method.

\begin{table*}[htbp]
\caption{Bootstraped mean system performance (best single-MOV values in bold) $^{\mathrm{a}}$}
\begin{center}
\setlength\tabcolsep{2pt}
\begin{tabular}{|c|c|c|c|c|c|c|}
\hline
\textbf{Objective}&\multicolumn{2}{|c|}{\textbf{Music Only}} & \multicolumn{2}{|c|}{\textbf{Speech Only}} & \multicolumn{2}{|c|}{\textbf{All Samples}}  \\
\cline{2-7} 
\textbf{Measure} & \textbf{\textit{$R_{pm}$}}& \textbf{\textit{$AES_m$}} & \textbf{\textit{$R_{pm}$}}& \textbf{\textit{$AES_m$}}  & \textbf{\textit{$R_{pm}$}}& \textbf{\textit{$AES_m$}} \\
\hline
\hline
PEAQ DI	&	0,55	&	2,96	&	0,77	&	2,6	&	0,72	&	2,56 \\
PEAQ $AvgLinDist_A$	&	0,84	&	1,8	&	0,81	&	2,4	&	\textbf{0,84}	&	\textbf{1,99} \\
PEAQ $RmsNoiseLoud_A$	&	0,56	&	2,9	&	\textbf{0,9}	&	\textbf{1,76}	&	0,8	&	2,23\\
PEAQ $RmsMissingComponents$ $^{\mathrm{b}}$	&	0,76	&	2,27	&	0,75	&	2,72	&	0,78	&	2,31\\
PEMO-Q PSMt	&	0,61	&	2,79	&	0,87	&	1,97	&	0,67	&	2,76\\
ViSQOLA NSIM	&	\textbf{0,86}	&	\textbf{1,79}	&	0,86	&	2,09	&	0,82	&	2,09\\
\hline
\hline
PEAQ $AvgLinDist_A$ + $RmsNoiseLoud_A$ + 	$RmsMissingComponents$ &	0,84	&	1,9	&	0,9	&	1,81	&	0,88	&	1,76\\
PEAQ ADVANCED RETRAIN	&	0,9	&	1,53	&	0,97	&	1	&	0,91	&	1,54\\
\hline
\multicolumn{4}{l}{$^{\mathrm{a}}$ All two-sided 95\% confidence intervals in the range of: $ R_p < \pm 0.01$, $AES < \pm 0.09$} \\
\multicolumn{4}{l}{$^{\mathrm{b}}$ Derived from $RmsNoiseLoudAsym_A$ \cite{PEAQ}}
\end{tabular}
\label{tab2}
\end{center}
\end{table*}

According to Table \ref{tab2}, ViSQOL Audio achieves a high and stable performance for all items overall. As the items in the database do not show complex temporal misalignments, the more complex (in comparison to PEAQ and PEMO-Q) alignment algorithm \cite{ViSQOLAudio} cannot be the cause of its high effectiveness. ViSQOLA NSIM presents the most balanced (and high) performance for all categories, although not the best in the category ``All Samples" for the evaluated data. The PEMO-Q similarity measure PSMt is among the best performing for speech items and among the worst performing for music items. It is also the worst performing degradation index overall for the ``All Samples" item category. 

The PEAQ DI quality measure still shows a low performance for all the combinations of sample types. However, PEAQ's $AvgLinDist_A$ shows the best performing measurements from all the single degradation indexes tested in the case of the sample type category ``All Samples" and the second best for ``Music Only". For music, $RmsNoiseLoud_A$ is the worst performing (except DI), but it is the best performing single degradation index for speech samples. These observations support the hypothesis that PEAQ still has an acceptable perceptual model for the quality estimation task despite the comparatively bad performance of PEAQ DI. Furthermore, results of Table \ref{tab2} show that added components and missing components play different roles in different signal types.

\begin{figure*}
\includegraphics[trim={0cm 0 0cm 0},clip,width=\textwidth, height=10cm]{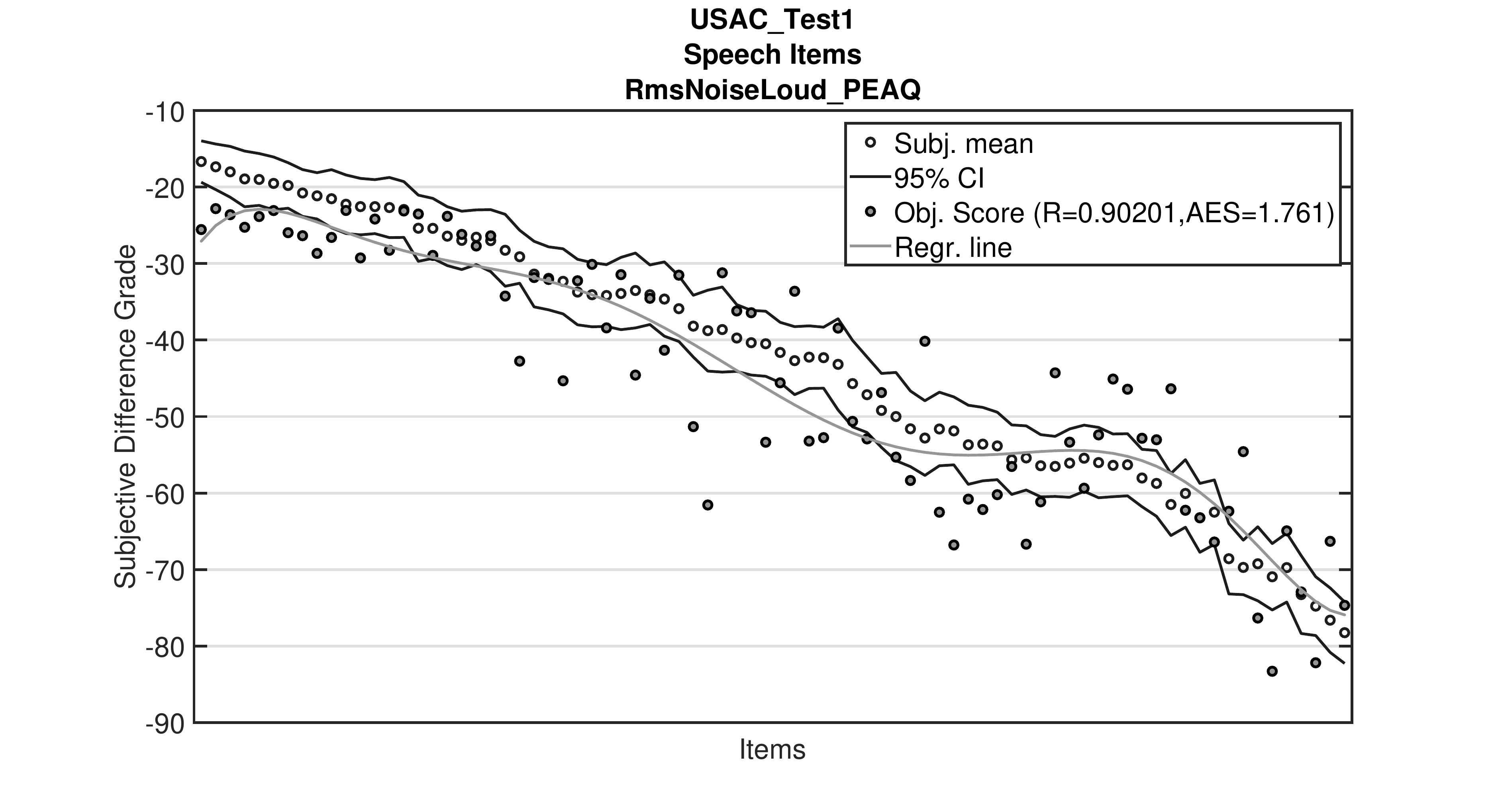}
\caption{Subjective difference grade scores  \cite{MUSHRA} (MUSHRA) and Monte Carlo average objective scores on test data (speech-only) after training with $RmsNoiseLoud_A$ MOV, which incorporates a quality model based on loudness of additive disturbances \cite{ThiedePEAQ}.}
\label{RmsNoiseLoud_Speech}
\end{figure*}

\begin{figure*}
\includegraphics[trim={0cm 0 0cm 0},clip,width=\textwidth, height=10cm]{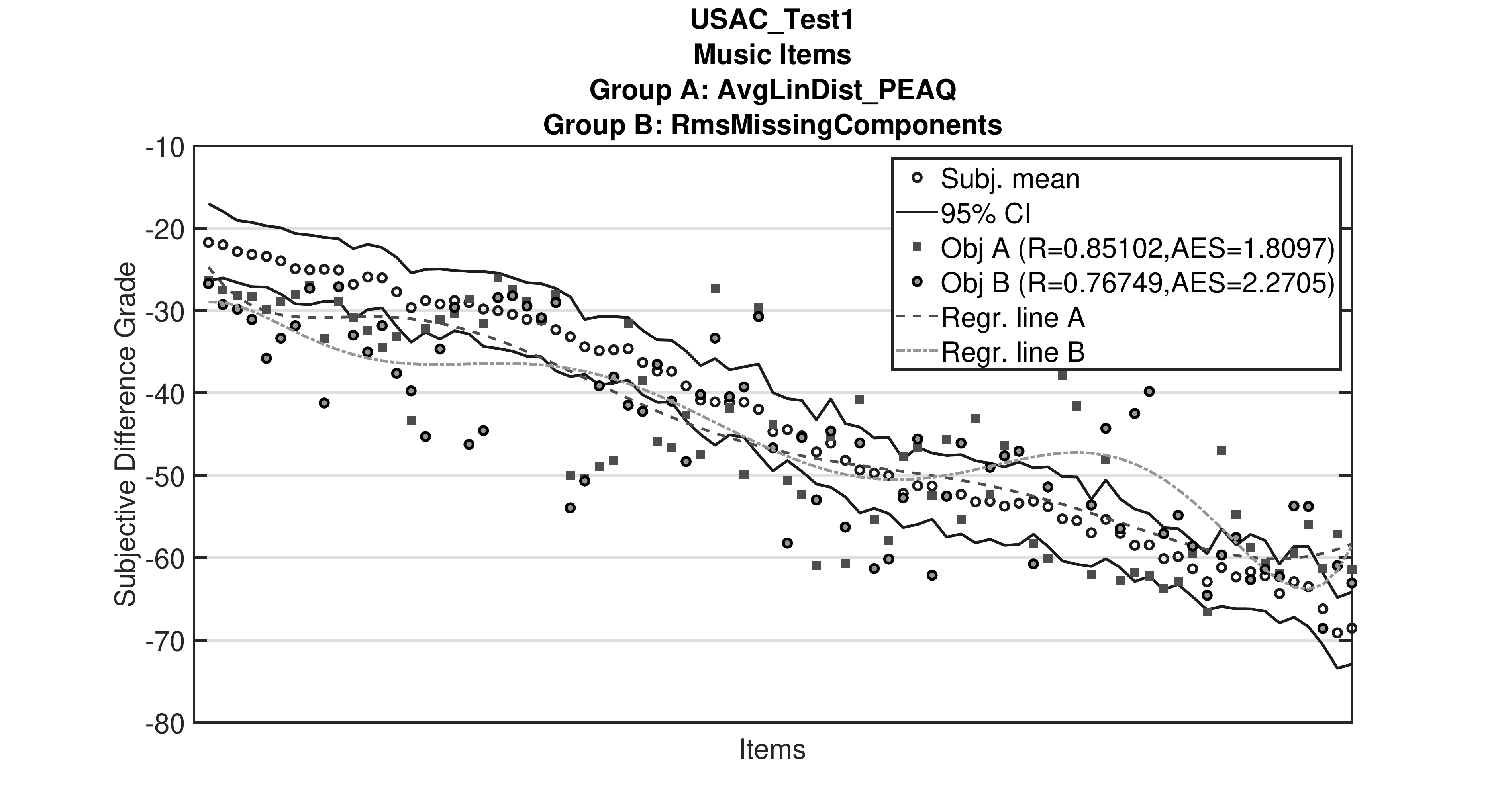}
\caption{Subjective difference grade scores \cite{MUSHRA} (MUSHRA score) and Monte Carlo average objective scores on test data (music-only) after training with $AvgLinDist_A$ and $RmsMissingComponents$ MOV.}
\label{fig2}
\end{figure*}

Further data visualization is provided in Figures \ref{RmsNoiseLoud_Speech} and \ref{fig2} showing SDG and mean ODG scores obtained using the three different disturbance loudness MOVs described in Section \ref{PEAQMOV} separated in speech and music-only items.  

Given the subjective database characteristics, the strong MOV-subjective score correlation associated with the partial loudness of disturbances model imply that this model is also suitable for a wider quality range -- namely the intermediate quality level as defined in \cite{MUSHRA} -- with a proper mapping. The lack of a proper remapping could explain the low discrimination power that PEAQ showed on lower bit rate codecs as reported in \cite{PoctaQuality}. However, from a visual examination of Fig \ref{RmsNoiseLoud_Speech} follows that the discrimination power of the additive disturbance loudness model for speech items is --on average-- in line with the subjective scores in all quality ranges. The observation that the PEAQ DI performance is low in all experiments -- but the performance of certain PEAQ MOVs is not -- suggests that a different MOV weighting in PEAQ is needed for an improved distortion index. Table \ref{tab2} shows that retraining PEAQ with the disturbance loudness-related MOVs greatly improves performance in comparison to PEAQ DI, which confirms that a more suitable mapping is beneficial. Remapping has also been reported to improve the performance POLQA for music content \cite{PoctaQuality}. Future work might consider including ``POLQA Music" --an objective quality system suitable for speech and audio -- in a similar analysis as the one presented in this work when it becomes available. The experimental data does not provide evidence that the underlying perceptual model of PEAQ would be the weak link in the processing chain, except maybe for the model of additive distortions in music items.

Although remapping greatly improves performance, measurements also show some possible limitations of the perceptual model. Considering the strong correlation values, it is clear that measuring the disturbance loudness of the added components represents a very meaningful descriptor of quality degradation in the case of speech signals, but is much less meaningful in the case of music signals. Besides retraining, further performance improvement could be achieved with the addition of a speech/music discriminator as part of the cognitive model, which in turn would allow the use of a suitable disturbance loudness analysis mode weighting (added versus missing components). PEMO-Q also weights added components more importantly than missing components on its PSMt measure \cite{PEMOQ}, which is in accordance with the data in Table \ref{tab2} showing a better performance of PSMt for speech-only items than music-only items. 

Analyzing loudness differences of missing components has proven a good strategy in the case of music signals, but not so much for speech. These results suggest that a better model for measuring the loudness of added components in music is needed. Recent studies indicate there might be different cognitive processes involved in the quality evaluation of speech and audio signals \cite{JASASpeechAudioCognitive}. An extended cognitive model considering these results in conjunction with the disturbance loudness model may be a promising update to PEAQ's current perceptual model. The role of an extended disturbance loudness model on the overall system performance could then be analyzed with respect to the contributions of a proper MOV-to-quality-score mapping.  

The $AvgLinDist_A$ MOV presents a comparatively high performance in the Monte Carlo simulations in all categories, although most of the signal degradations in the database are not due to linear distortions. By definition, linear distortions do not introduce additional frequency components. Therefore, part of the performance could be explained by the fact that the MOV also performs an indirect analysis of missing components. However, its superior performance in comparison to the actual measurement of missing components hints to some additional benefits (compare performance of $AvgLinDist_A$ and $RmsMissingComponents$ in Fig. \ref{fig2}). Further analysis of Fig \ref{fig2} shows a more stable performance of $AvgLinDist_A$ than $RmsMissingComponents$ in the higher and lower quality extremes (smaller errors). The data suggest that a deeper analysis of the inner workings in the calculation of the $AvgLinDist_A$ might shed some light on its performance, especially concerning the level and pattern adaptation stages, as described in \cite{ThiedePEAQ}.

The best performance in Table \ref{tab2} is given by retraining all five MOVs from PEAQ Advanced. Accordingly, future work might include a similar analysis as the one presented in this work for the two remaining MOVs.

\section{Summary and Conclusion}

This work used Monte Carlo simulations to compare some of the most used objective quality measurement systems of perceptually coded audio signals while keeping focus on their underlying perceptual model rather than the quality scale mappings.
Individual MOVs showed an acceptable performance whereas mapping-dependent distortion measures did not. The data supports the finding that PEAQ's perceptual model -- particularly the modeling of disturbance loudness -- is still appropriate for newer audio codecs, even in the intermediate quality range. Consequently, the remapping of the perceptual model's MOV to a single quality score using training data from listening tests performed over newer audio codecs can improve the overall performance of PEAQ. 

The disturbance loudness model showed a significant difference in performance depending on the sample type category evaluated. Some of PEAQ's MOVs performances are on par with other newer systems. Further performance for the joint evaluation the quality of speech and audio signals in PEAQ could be improved, provided that the correct disturbance loudness analysis mode is used for each case. An improved cognitive model applied to the pattern adaptation stage could also improve PEAQ on this matter.

\bibliographystyle{IEEEtran}
\bibliography{papers_st_polqa}

\begin{thebibliography}{10}
\providecommand{\url}[1]{#1}
\csname url@samestyle\endcsname
\providecommand{\newblock}{\relax}
\providecommand{\bibinfo}[2]{#2}
\providecommand{\BIBentrySTDinterwordspacing}{\spaceskip=0pt\relax}
\providecommand{\BIBentryALTinterwordstretchfactor}{4}
\providecommand{\BIBentryALTinterwordspacing}{\spaceskip=\fontdimen2\font plus
\BIBentryALTinterwordstretchfactor\fontdimen3\font minus
  \fontdimen4\font\relax}
\providecommand{\BIBforeignlanguage}[2]{{%
\expandafter\ifx\csname l@#1\endcsname\relax
\typeout{** WARNING: IEEEtran.bst: No hyphenation pattern has been}%
\typeout{** loaded for the language `#1'. Using the pattern for}%
\typeout{** the default language instead.}%
\else
\language=\csname l@#1\endcsname
\fi
#2}}
\providecommand{\BIBdecl}{\relax}
\BIBdecl

\bibitem{PEAQ}
{\relax ITU-R Rec. BS.1387}, \emph{Method for objective measurements of
  perceived audio quality}, Geneva, Switzerland, 2001.

\bibitem{PoctaQuality}
P.~{Počta} and J.~G. {Beerends}, ``Subjective and objective assessment of
  perceived audio quality of current digital audio broadcasting systems and
  web-casting applications,'' \emph{IEEE Transactions on Broadcasting},
  vol.~61, no.~3, pp. 407--415, Sep. 2015.

\bibitem{ThiedePEAQ}
T.~Thiede, W.~C. Treurniet, R.~Bitto, C.~Schmidmer, T.~Sporer, J.~G. Beerends,
  and C.~Colomes, ``{PEAQ} - the {ITU} standard for objective measurement of
  perceived audio quality,'' \emph{J. Audio Eng. Soc.}, vol.~48, no. 1/2, pp.
  3--29, January/February 2000.

\bibitem{beerends1992a}
J.~G. Beerends and J.~A. Stemerdink, ``A perceptual audio quality measure based
  on a psychoacoustic sound representation,'' \emph{J. Audio Eng. Soc},
  vol.~40, no.~12, pp. 963--978, 1992.

\bibitem{BS1116}
{\relax ITU-R Rec. BS.1116}, \emph{{Methods for the subjective assessment of
  small impairments in audio systems}}, Geneva, Switzerland, 2015.

\bibitem{PEMOQ}
R.~{Huber} and B.~{Kollmeier}, ``{PEMO-Q}—a new method for objective audio
  quality assessment using a model of auditory perception,'' \emph{IEEE
  Transactions on Audio, Speech, and Language Processing}, vol.~14, no.~6, pp.
  1902--1911, Nov 2006.

\bibitem{kmpf2010standardization}
S.~Kämpf, J.~Liebetrau, S.~Schneider, and T.~Sporer, ``{Standardization of
  PEAQ-MC: Extension of ITU-R BS.1387-1 to Multichannel Audio},'' in
  \emph{Audio Engineering Society Conference: 40th International Conference:
  Spatial Audio: Sense the Sound of Space}, Tokyo, Oct 2010.

\bibitem{POLQAcite}
{\relax ITU-T Rec. P.863}, \emph{Perceptual Objective Listening Quality
  Assessment}, Geneva, Switzerland, 2014.

\bibitem{ViSQOLAudio}
C.~Sloan, N.~Harte, D.~Kelly, A.~C. Kokaram, and A.~Hines, ``Objective
  assessment of perceptual audio quality using {ViSQOLAudio},'' \emph{IEEE
  Transactions on Broadcasting}, vol.~PP, no.~99, pp. 1--13, 2017.

\bibitem{torcoli2018comparing}
M.~Torcoli and S.~Dick, ``Comparing the effect of audio coding artifacts on
  objective quality measures and on subjective ratings,'' in \emph{Audio
  Engineering Society Convention 144}, May 2018.

\bibitem{USACCodec_short}
M.~Neuendorf \emph{et~al.}, ``The iso/mpeg unified speech and audio coding
  standard—consistent high quality for all content types and at all bit
  rates,'' \emph{J. Audio Eng. Soc}, vol.~61, no.~12, pp. 956--977, 2013.

\bibitem{USACdatabase}
\BIBentryALTinterwordspacing
{\relax ISO/IEC JTC1/SC29/WG11}, ``{USAC} verification test report {N12232},''
  International Organisation for Standardisation, Tech. Rep., 2011. [Online].
  Available:
  \url{{https://mpeg.chiariglione.org/standards/mpeg-d/unified-speech-and-audio-coding/unified-speech-and-audio-coding-verification-test}}
\BIBentrySTDinterwordspacing

\bibitem{MUSHRA}
{\relax ITU-R Rec. BS.1534}, \emph{{Method for the subjective assessment of
  intermediate quality levels of coding systems}}, Geneva, Switzerland, 2015.

\bibitem{Jekabsons_areslab}
G.~Jekabsons, ``Areslab: Adaptive regression splines toolbox for matlab.
  http://www.cs.rtu. lv/jekabsons/regression.html,'' 2019.

\bibitem{PEASS}
V.~{Emiya}, E.~{Vincent}, N.~{Harlander}, and V.~{Hohmann}, ``Subjective and
  objective quality assessment of audio source separation,'' \emph{IEEE
  Transactions on Audio, Speech, and Language Processing}, vol.~19, no.~7, pp.
  2046--2057, Sep. 2011.

\bibitem{Visqol_soft}
A.~Hines, E.~Gillen, D.~Kelly, J.~Skoglund, A.~Kokaram, and N.~Harte, ``Visqol
  audio matlab implementation. http://www.sigmedia.tv/tools,'' Accessed 2019.

\bibitem{JASASpeechAudioCognitive}
\BIBentryALTinterwordspacing
R.~Huber, S.~Rählmann, T.~Bisitz, M.~Meis, S.~Steinhauser, and H.~Meister,
  ``Influence of working memory and attention on sound-quality ratings,''
  \emph{The Journal of the Acoustical Society of America}, vol. 145, no.~3, pp.
  1283--1292, 2019. [Online]. Available:
  \url{https://doi.org/10.1121/1.5092808}
\BIBentrySTDinterwordspacing

\end{thebibliography}

\end{document}